\documentclass[aps,prl,showpacs,twocolumn,groupedaddress,amssymb]{revtex4}

\usepackage{graphicx}
\usepackage{dcolumn}
\usepackage{bm}

\begin{document}

\title{Relativistic electron beam, Backward Raman scattering and soft X-ray laser}

\author{S. Son}
\affiliation{169 Snowden Lane, Princeton, NJ, 08540}

\begin{abstract}
A scheme for soft x-ray lasers  is proposed. The backward Raman scattering between an intense visible-light laser and a relativistic electron beam results in soft x-ray light via the Doppler shift. One of the most intense soft x-ray light sources is contemplated.  
\end{abstract}


\maketitle

There have been exploding interests in a \textit{high-intensity} (\textit{high-power})   x-ray light source,  which would  
 be critical  in the atomic spectroscopy, the dynamical imaging of biological processes, the next-generation semi-conductor lithography and many others. 
While its impacts on the sciences and the practical applications are immense, 
the most advanced  x-ray sources~\cite{sonttera, colson, songamma,Gallardo, sonltera,freelaser, freelaser2, Free, Free2} are not intense or efficient enough  for mentioned applications.





One emerging way to generate intense x-ray  is  to utilize the scattering between an intense laser and a relativistic electron beam.
 This methods becomes very attractive as the table-top electron beam accelerator (intense laser) is readily available.
In this paper, the author proposes a soft x-ray light source by further amplifying the interaction between the laser and the electron beam via the backward Raman scattering (BRS).
There have been some efforts to generate the x-ray via the BRS~\cite{brs2, brs3, drake, brs}, where
 the second or third harmonic interaction~\cite{brs,  brs2, brs3} or the milimeter wave based the conventional magnets~\cite{brs} is utilized to generate XUV light. 
 However, its efficency (intensity and power) was low and 
the wave-length of the resulted XUV light was comparable to 100 nm, that is  too large for many important applications. 
On the other hand, what has never been attempted  is to utilize  the relativistic Doppler effect directly or the first harmonic of the BRS.
Noting that the  great advances in intense visible-light lasers~\cite{cpa, cpa2, cpa4} and  dense relativistic electron beams~\cite{monoelectron, ebeam} have been made recently, 
the possibility to   directly use the first harmonic and the relativistic Doppler's effect is  proposed in this paper.  

\begin{figure}
\scalebox{0.3}{
\includegraphics{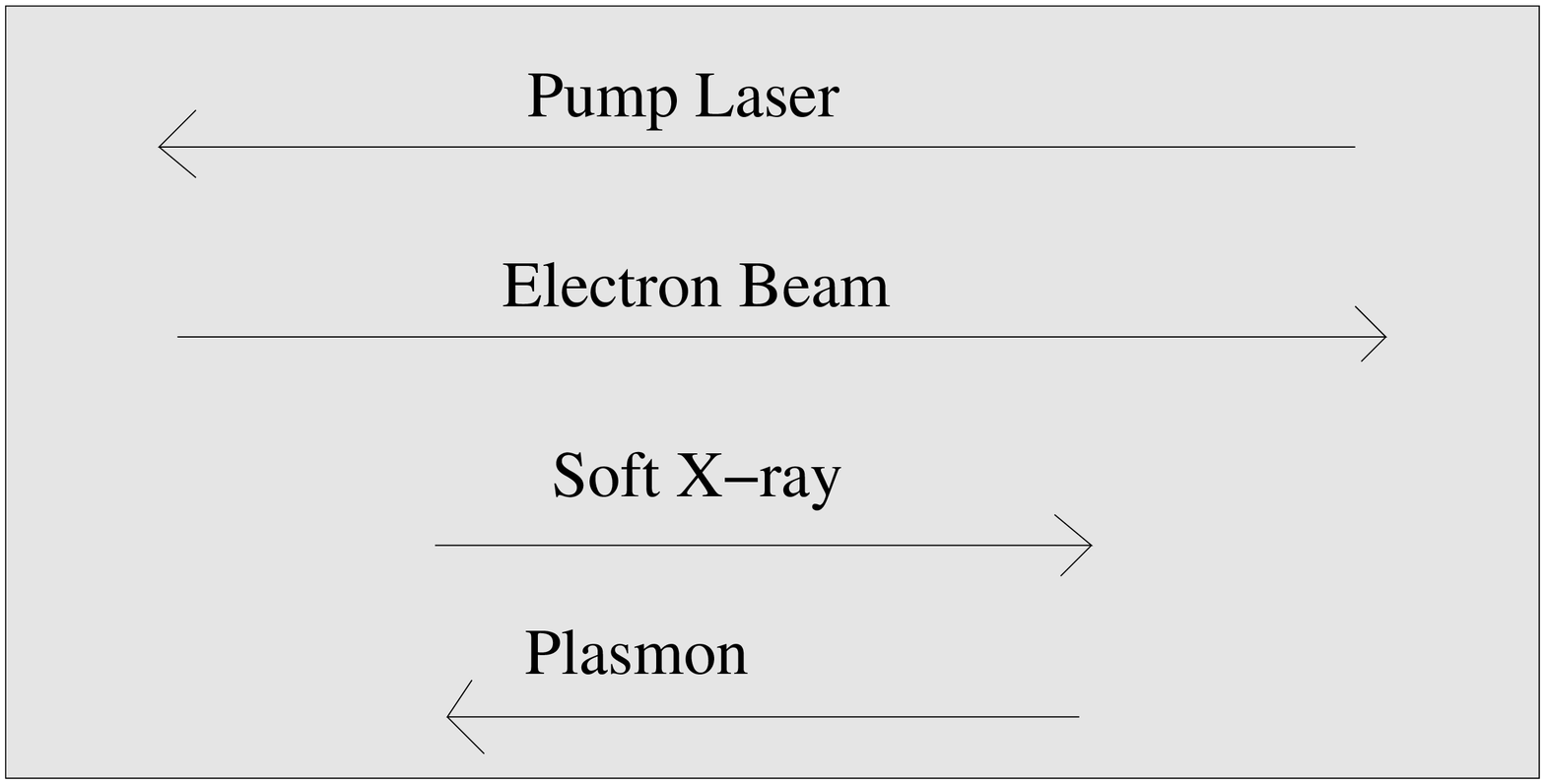}}
\caption{\label{fig:1}
The diagram  about the propagation direction of the pump laser, the plamson, the soft x-ray and 
 the electron beam. 
}
\end{figure}

The main idea is  as follows. First, a dense electron beam and an intense visible-light laser (pump laser) propagate in the \textit{opposite direction}. 
Second, 
the ponderomotive interaction between the laser and a weak or background x-ray light (seed pulse)  excites a Langmuir wave. Third, 
 the backward Raman scattering between the Langmuir wave and the laser 
 amplifies  the x-ray light (seed pulse)  in the \textit{electron  beam direction} and up-shfited in frequency via the relativistic Doppler's effect. See Fig.~(\ref{fig:1}) for the direction of the visible-light laser, the soft x-ray light (seed pulse) and the Langmuir wave. 
In this paper,  the physics of the above processes will be analyzied and the prospect  as a x-ray light source will be presented. 
The analysis here suggests the most efficient and intense x-ray light source with  
the achieved x-ray  wave length of 10 nm rather than 100 nm, which is ideal for many important applications. 


To begin, let me 
consider an intense visible-laser with frequency $\omega_{p0} $ (wave vector $k_{p0}$) and a counter-propagating dense electron beam with the electron density $n_0$ and the relativistic factor $\gamma_0= (1-v_0^2/c^2)^{-1/2}$, where $v_0$ is the velocity of the electron beam. In the co-moving frame with the electron beam, the electron density decreases to $n_1 = n_0 / \gamma_0$ due to the length dilation. The 1-D BRS three-wave interaction in the \textit{co-moving} frame between the pump, the seed and a Langmuir wave is described by~\cite{McKinstrie}:
\begin{eqnarray}
\left( \frac{\partial }{\partial t} + v_p \frac{\partial}{\partial x} + \nu_1\right)A_p  = -ic_p A_s A_3  \nonumber \mathrm{,}\\
\left( \frac{\partial }{\partial t} + v_s \frac{\partial}{\partial x} + \nu_2\right)A_s  = -ic_s A_p A^*_3   \label{eq:2} \mathrm{,} \\
\left( \frac{\partial }{\partial t} + v_3 \frac{\partial}{\partial x} + \nu_3\right)A_3  = -ic_3 A_p A^*_s  
\nonumber \mathrm{,}
\end{eqnarray}
where $A_i= eE_{i1}/m_e\omega_{i1}c$ is the ratio of the electron quiver velocity of the pump pulse ($i=p$) and the seed pulse ($i=s$) relative to the velocity of the light $c$, $E_{i1}$ is the electric field of the E\&M pulse, $A_3 = \delta n_1/n_1$ is the Langmuir wave amplitude, $\nu_1 $ ($\nu_2$) is the rate of the inverse bremsstrahlung of the pump (seed), $\nu_3$ is the plasmon decay rate, $ c_i = \omega_3^2/ 2 \omega_{i1}$ for $i=p, s$, $c_3 = (ck_3)^2/2\omega_3$, $ \omega_{p1} $ ($ \omega_{s1} $) is the frequency of the pump (seed) pulse and $\omega_{3} \cong \omega_{pe} / \sqrt{\gamma_0} $ is the plasmon  wave frequency. 

In the co-moving frame, the wave vector satisfies the usual dispersion relationship, $\omega_1^2 =   \omega_{pe}^2/\gamma_0+ c^2 k_1^2$, where $\omega_1$ ($k_1$) is the wave frequency (vector) and $\omega_{pe}^2 = 4 \pi n_0 e^2 /m_e$ is the plasmon frequency. Denote  the wave vector  (the corresponding wave frequency) of the pump laser (the seed pulse or soft x-ray) in the co-moving frame as $k_{p1}$, $k_{s1} $, $\omega_{p1} $ and $\omega_{s1} $, and the laboratory-frame counterparts as $k_{p0}$, $k_{s0} $, $\omega_{p0} $ and $\omega_{s0} $. The Lorentz transform prescribes the following relationship: 

\begin{eqnarray} 
\omega_{p0} &=& \gamma_0 \left[ \sqrt{\omega_{pe}^2/\gamma_0 + c^2 k_{p1}^2 } - vk_{p1} \right] \mathrm{,}  \label{eq:lorentz1} \\  \nonumber \\
k_{p0} &=&  \gamma_0 \left[ k_{p1} - \frac{\omega_{p1} }{c}  \frac{v_0}{c} \right] \mathrm{,} \label{eq:lorentz2} \\ \nonumber \\ 
 \omega_{s0} &=& \gamma_0 \left[ \sqrt{\omega_{pe}^2/\gamma_0 + c^2 k_{s1}^2 } + 
vk_{s1} \right] \mathrm{,}  \label{eq:lorentz3} \\  \nonumber \\ 
k_{s0} &=&  \gamma_0 \left[ k_{s1} + \frac{\omega_{s1} }{c}  \frac{v_0}{c} \right]\mathrm{.} \label{eq:lorentz4} \\ \nonumber 
\end{eqnarray}
Using Eqs.~(\ref{eq:lorentz1}), (\ref{eq:lorentz2}), (\ref{eq:lorentz3}) and  (\ref{eq:lorentz4}), the pump laser (seed pulse or soft x-ray) can be transformed from the co-moving frame to the laboratory frame or vice versa. 

The energy and momentum conservation of Eq.~(\ref{eq:2}) lead to  
\begin{eqnarray} 
 \omega_{p1} &=& \omega_{s1} + \omega_{3}  \nonumber 
\mathrm{,} \nonumber \\  
k_{p1} &=& k_{s1} +k_3\mathrm{,} \label{eq:cons}  
\end{eqnarray}
where $k_3$ is the plasmon wave vector. For a given pump frequency $\omega_{p0} $, $k_{p1} $ ($\omega_{p1} $) is obtained from  Eq.~(\ref{eq:lorentz1}), $k_{s1} $ ($\omega_{s1}$) is from Eq.~(\ref{eq:cons}) and, finally,  $k_{s0} $ ($\omega_{s0}$) is from Eqs.~(\ref{eq:lorentz3}) and (\ref{eq:lorentz4}). In the limit when $ck_{s1} \gg \omega_3 $, $\omega_{s0} \cong 2 \gamma_0 (\omega_{p1} - \omega_3)$ or  

\begin{equation} 
\omega_{s0} \cong 4\gamma_0^2 \left[ \omega_{p0} -  \frac{\omega_{pe}}{2 (\gamma_0)^{-3/2}}\right] \mathrm{,}\label{eq:down}
\end{equation} 
using  $\omega_{p1} \cong 2 \gamma_0  \omega_{p0}$ and $\omega_3 \cong \omega_{pe} /\sqrt{\gamma_0} $. The Eq.~\ref{eq:down} describes the frequency up-shift of the pump pulse into the soft x-ray by the relativistic Doppler's effect.

The growth rate of the BRS is obtained from the linear analysis of Eq.~(\ref{eq:2}). When $|A_p| \gg |A_s|$  and $|A_p| \gg |A_3|$, the linearization of Eq.~(\ref{eq:2}), in the form $A_{i} = A_{\omega,i}\exp (i \omega t) $ with $i= s$ or $i= 3$, leads to  

\begin{eqnarray} 
\omega^2 &+& (\nu_3 + \nu_2) \omega + (\nu_3 \nu_2) - c_s c_3 |A_{p}|^2=0  \mathrm{,} 
\label{eq:inst} 
\end{eqnarray}  
If $\nu_2 \cong 0$ and $\nu_3 \ll ( c k_{p1} \omega_3)^{1/2} |A_{p}|$, the growth rate, the imaginary part of the solution of Eq.~(\ref{eq:inst}), is $\Gamma_1 \cong  (c_s c_3)^{1/2} |A_{p}| $. In the limiting case when $ck_{p} \gg \omega_3 $, it can be simplified to 

\begin{equation} 
\Gamma_1 \cong \left(\sqrt{\frac{2\omega_{pe} \omega_{p0}}{\gamma_0^{1/2}} }
\right) |A_{p}|  
\label{eq:inst2} \mathrm{,}
\end{equation}
where  $k_{p1} \cong k_{s1} \cong 2 \gamma_0 k_{p0} $, $\omega_3 = \omega_{pe} / \gamma_0 $, $E_{p1} / E_{p0} \cong 2 \gamma_0$, $\omega_{p0} / \omega_{p1} \cong 1/  2 \gamma_0$ and thus   $A_{p}= eE_{p1}/m_e\omega_{p1} \cong  eE_{p0}/m_e\omega_{p0}$. If we define the electron beam length as $L_b$ and the laser length as $L_l$, the beam length increases to $\gamma_0 L_b $ and the laser length decreases to $L_l /  2 \gamma_0 $ in the co-moving frame so that the interaction time between the beam and laser is  $ \tau \cong \min(\gamma_0 L_b/c, L_l/2\gamma_0/c) $. The gain-per-length $g$  is then 
\begin{eqnarray} 
 g &=& \Gamma_1 \tau / L_b = \Gamma_1 \gamma_0 /c \ \ \  \mathrm{if} \  L_b < 2 \gamma_0^{-2} L_l \mathrm{,} \nonumber \\ 
 g &=& \Gamma_1 \tau / L_l = \Gamma_1 /2\gamma_0 c \ \   \mathrm{otherwise.} 
 \label{eq:gain}  
\end{eqnarray}
Considering an electron beam with the electron density of $n_0 = 10^{20}/ \mathrm{cc} $ and ND:YAG laser with the wave length of 1 $\mu \mathrm{m} $, 
 $A_p  \cong 0.03 $ and $\Gamma_1 \cong 0.4 \times 10^{14} / \sec $ from Eq.~(\ref{eq:inst2}) if the laser intensity is  $10^{15} \ \mathrm{W} / \mathrm{cm}^2 $. If $\gamma_0 \cong 10$, the amplified wave has the wave length of 2.5 nm from Eq.~(\ref{eq:down}) and the gain-per-length is, from Eq.~(\ref{eq:gain}), $ g =  0.13 \times 10^{5} / \mathrm{cm} $ when $L_b < 2 \gamma_0^{-2} L_l$ and $ g =  0.06 \times 10^{3} / \mathrm{cm} $ when   $L_b > 2 \gamma_0^{-2} L_l$. For the same condition but with the laser intensity of $10^{18} \ \mathrm{W} / \mathrm{cm}^2 $,    $\Gamma_1 \cong 1.5 \times 10^{15} / \sec $ and the gain-per-length is $ g =  0.5 \times 10^{6} / \mathrm{cm} $ when $L_b < 2 \gamma_0^{-2} L_l$ and $ g =  0.25 \times 10^{4} / \mathrm{cm} $ when   $L_b > 2 \gamma_0^{-2} L_l$; 
the gain-per-length is   higher than any other conventional soft x-ray lasers by a few factors~\cite{laser, laser1, gain}

\begin{table}[t]
\centering
\begin{tabular}{|c||cccccccc||}
	\hline
  Type &  $\mathbf{\lambda_{s0}} $   &   $I_{15}$   &  $\gamma_0$ &  $A_1$ & $\Gamma_{13}$ & $\mathbf{g_1}$ & $g_2$ &  $n_c$   \\
	\hline \hline 

N    & \textbf{2.5} &     1 &   10&      0.02&       5.18&   \textbf{17}&       0.08&       64 \\
N  &   \textbf{2.5} &     100 &     10&      0.6&       164&    \textbf{547} &      2.7&        64 \\ 
N  &   \textbf{2.5} &     0.01&     10 &     0.002&        0.5&      \textbf{1.7}&       0.009&       64\\
N  &   \textbf{10} &    1 &  5 &       0.02&        4.3&      \textbf{7.2}&      0.14&       4 \\
N  &   \textbf{10} &   $10^{3} $ &     5 &       0.6 &        137 &     \textbf{230} &     4.6 &        4 \\
N  &    \textbf{27.8}&    1       &  3  &      0.02 &        3.8 &    \textbf{3.8} &      0.21 &      0.51 \\ 
N  &   \textbf{27.8}&    0.01    &  3   &     0.002 &       0.38 &     \textbf{0.38}&       0.02 &        0.51 \\ 
N  &   \textbf{27.8}&    $10^{3}$ &   3  &      0.6&        121 &    \textbf{121} &      6.7 &      0.51 \\
C  &   \textbf{2.5} &     1     &   31.6  &   0.2  &       21.8 &   \textbf{30.5} &       0.1 &       6.4 \\
C  &   \textbf{2.5} &     30    &   31.6  &  1.1   &      119.7 &  \textbf{1262}  &     0.6   &    6.4 \\ 
C  &   \textbf{10}  &     1     &  15.8   &  0.2 &      18.3    &  \textbf{96.9}  &     0.2   &     0.4 \\ 
C  &   \textbf{10}  &     0.001 &   15.8  &   0.006 &       0.5 &     \textbf{3}  &     0.006 &      0.4 \\
C  &   \textbf{25}  &     1     & 10      &  0.2 &         16.4 &    \textbf{54.7}&        0.27 &      0.064 \\ 
C  &   \textbf{25}  &     0.01  &  10     &  0.02&        1.64  &    \textbf{5.5} &       0.027 &      0.064 \\
\hline
\end{tabular}
\caption{The laser and electron beam parameters and the characteristic of the soft x-ray  radiation \label{tb}. In this example, I assume that $n_0 = 10^{20} \ / \mathrm{cc}$. In the table, N (C) stands for the NA:YAG laser (CO2 laser) with the wave length of $1 \ \mu \mathrm{m} $ ($10 \ \mu \mathrm{m} $), $I_{15}$ is the laser intensity normalized by $10^{15} \ \mathrm{W} / \mathrm{cm}^2$, $\gamma_0$ is the relativistic factor, $A_1$ is the quiver velocity divided by the velocity of light as defined in Eq.~(\ref{eq:2}),  $\lambda_{s0} $ is the wave length of  the seed pulse obtained from Eq.~(\ref{eq:down}) in the unit of  1 nm, $\Gamma_{13} $ is the growth rate normalized by $10^{13} / \sec$ as given in Eq.~(\ref{eq:inst2}),  $g_1$ ($g_2$) is the gain-per-length from Eq.~(\ref{eq:gain}) in the unit of $10^3 \ /\mathrm{cm} $ and $n_c$ is the lower bound of the density given in Eq.~(\ref{eq:cond}) normalized by $10^{16}\  /\mathrm{cc} $. 
}
\end{table}

The estimation of the conversion efficiency from the pump laser energy to the seed pulse follows. Define $\epsilon_1 $ as the conversion efficiency in the co-moving frame. Let $\mathrm{E}_{p0} $ be  the total energy of the pump laser in the laboratory frame. Then,  the pump energy in the co-moving frame is $\mathrm{E}_{p1} \cong 2 \gamma_0 \mathrm{E}_{p0}$ via the Doppler effect, the energy converted to the seed pulse is $\mathrm{E}_{s1} \cong  \epsilon_1 \mathrm{E}_{p1} = 2 \gamma_0 \epsilon_1 \mathrm{E}_{p0} $, and the total energy of the seed pulse in the laboratory frame is $\mathrm{E}_{s0} \cong 2 \gamma_0 \mathrm{E}_{s1} \cong 4 \gamma_0^2 \epsilon_1 \mathrm{E}_{p0} $, resulting the effective conversion efficiency as
\begin{equation}
\epsilon_0 = 4 \gamma_0^2 \epsilon_1 \mathrm{.} \label{eq:eff} 
\end{equation}
Given the fact that $\epsilon_1$ can be comparable to the unit in an optimistically envisioned scenario~\cite{Fisch, malkin1, sonbackward, BBRS, BBRS2, BBRS3}, it is possible that $\epsilon_0 = 4 \gamma_0^2 \epsilon_1 > 1 $ or the conversion efficiency can be larger than 100 percents. This extraordinary high conversion efficiency can be explained as follows. In the co-moving frame, the laser will be scattered into a seed photon and a plasmon, during which   the plasmon (the seed photon) acquires the momentum $-k_{s1} - k_{p1} $ ($k_{s1} $). In the laboratory frame, the momentum (energy) of the plasmon is much larger than that of the pump photon due to the Doppler effect and the kinetic energy loss of the electron beam, when the plasmon is excited, is  even higher than the pump photon energy. In other words, the seed pulse gets most of its energy not from the pump pulse but from the electron beam (by the ratio of $1:4\gamma_0^2$).  The relativistic electron beam can support the wave with negative energy in the laboratory frame.  

In the high conversion efficiency as discussed, the BRS might be already in the non-linear saturated regime, in which $|A_{s}| $ is  as intense as $|A_{p}|$.  The wave breaking limit  might be simply put as $A_3 < 1$.  From  Eq.~(\ref{eq:2}), 
\begin{equation}
 A_3 \cong |A_p A_s| \left( \frac{c^2 k_{p1}^2}{ \omega_3 \nu_3}\right) < 1 
\mathrm{,} \label{eq:sat}
\end{equation} 
where I assume that $k_{p1} \cong k_{s1}$. Using Eq.~(\ref{eq:2}) and Eq.~(\ref{eq:sat}), the growth rate of the seed pulse is  
$\Gamma_1 \cong (\omega_3^2 / ck_{s1})( c^2 k_{p1}^2 /\omega_3\nu_3)|A_p|^2$. Assuming $|A_p| \cong |A_s|$,  
\begin{equation} 
\Gamma_1 = \left(\frac{\omega_3^2}{ ck_{s1}} A_3\right) < \left(\frac{\omega_{pe}^2}{  ck_{s1} \gamma_0} \right)\label{eq:upper} \mathrm{,}
\end{equation} 
which prescribes the upper-bound of the growth rate in the non-linear saturated regime.

In order for the current scheme to be plausible, there are two necessary conditions to be met for the laser beam intensity and electron beam density. One necessary condition  would be that 
 $\Gamma_1 \tau >1 $ for a sufficient amplification or
\begin{equation} 
A_{p1} > 
\sqrt{\left(\frac{1}{\tau^2}\right)\left(\frac{ c k_{s1} \gamma_0 }{c^2 (k_{p1} + k_{s1})^2 \omega_{pe} }\right)} \mathrm{,} \label{eq:cond1}
\end{equation}
which can be satisfied by currently available intense visible-light lasers~\cite{cpa, cpa2, cpa4}. Another necessary condition for the collective BRS is $n_0 / \gamma_0 >  (k_{p1} + k_{s1})^3 / (2 \pi)^3 $, as only the Langmuir wave with $n_0 /\gamma_0 \gg (k_3/2\pi)^2 $ is 
 a collective wave. In the limiting case when $ c k_{p1} \gg \omega_3$,  it is simplified to 
\begin{equation} 
 n_0 >    64 \times \gamma_0^4 \left(\frac{k_{p0}^3}{  8\pi^3 }  \right)
 \cong \left(\frac{1}{ \gamma_0^2} \right) \left(\frac{k_{s0}^3}{  8\pi^3 }  \right)
   \label{eq:cond}  \mathrm{,}
\end{equation} 
where  $k_{p1} \cong k_{s1} \cong 2 \gamma_0 k_{p0}$ and $k_{s0} \cong 4 \gamma_0^2 k_{p0}$ are assumed. In addition,  the wave vector of the Langmuir wave should be larger than the Debye length

\begin{equation}
k_3 < 0.5 \times \lambda_{de}^{-1}   \label{eq:cond2}
\end{equation} 
where $\lambda_{de}^{-2} =  4 \pi n_0 e^2 /T_e \gamma_0 m_e $. The condition given by Eqs.~(\ref{eq:cond}) and (\ref{eq:cond2}) is the minimum lower bound of the electron beam density for the BRS compression and can be met easily in the currently available dense electron beams~\cite{monoelectron, ebeam} while the electron temperature low enough to satisfy Eq.~(\ref{eq:cond2}).

\begin{figure}
\scalebox{1.0}{
\includegraphics[width=0.7\columnwidth, angle=270]{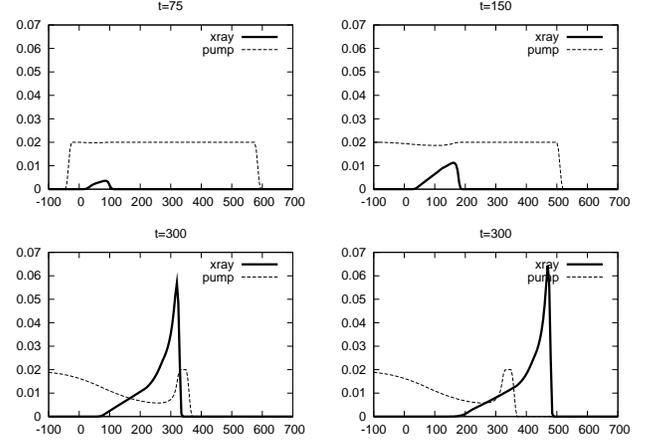}}
\caption{\label{fig:2}
The 1-D simulation of the BRS in the co-moving frame,  where $n_0 = 10^{20} \ \mathrm{cm^{-3}} $ and  $\gamma_0=3.0$. The $x$-axis is normalized by $ x_c = \omega_{\mathrm{pe}}/c\sqrt{\gamma_0} $ ($x_c \cong 0.7 \ \mu \mathrm{m}$) and the y-axis is the quiver velocity divided by the velocity of the light ($\mathrm{A} = A_p$ or $\mathrm{A}= A_s$), as defined in Eq.~(\ref{eq:2}). 
Initially ($t=0$), the x-ray pulse is located at $x=0$ with the peak of $\mathrm{A} = 0.0001$ corresponding to 
$I_{15} = 1.5 \times 10^{11} \ \mathrm{W} / \mathrm{cm}^2 $.  The inital pump laser  is located between $x=0 $ and $x=600 \  x_c $, with  the peak of $\mathrm{A} = 0.02$ corresponding to $I= 10^{15} \ \mathrm{W} / \mathrm{cm}^2 $. 
The pump (x-ray) propagetes from the right (left) to the left (right).
The top-left is the x-ray pulse and pump at  $ t= 75  \ \tau_c$ where $\tau_c = 2.5 $ femto-seconds. 
   The top-right is the x-ray pulse and the pump at $t=150 \ \tau_c$.   The bottom-left is the x-ray pulse and the pump at $t=225 \ \tau_c$. The bottom-right is the x-ray pulse and the pump at $t=300 \ \tau_c$. 
The final peak intensity of the x-ray, which is proportional to $\mathrm{A}^2$, is larger than the initial intenisty  by a factor of $3.6 \times 10^{5}$.
}
\end{figure}

To be more specific, the gain analysis for various electron beams and lasers are provided in the Table~(\ref{tb}). Those results suggest that the scheme is plausible for a wide range of frequencies, the requirement of the laser intensity is moderate compared to other laser based schemes~\cite{songamma, sonttera,  sonltera,  Gallardo, brs, brs2, brs3}, the gain-per-length can be as high as $10^6 / \mathrm{cm} $ and the optimal relativistic factor is  $ 3\le \gamma_0 \le 30$. In Table~(\ref{tb}), for the cases of the CO2 laser with the $\gamma_0 = 31.6 $ and $\gamma_0=15.8$, the temperature constraint given by Eq.~(\ref{eq:cond2}), assuming the electron density is $n_c$, might be too stringent $T_e < 10  \ \mathrm{eV} $. Generally, the generation of the short wave-length x-ray by the CO2 laser could put the density requirement higher than given by $n_c$ for reasonable electron temperature. If $I_{15} \geq 10^3 $ for the ND:YAG laser as in the Table~(\ref{tb}), the electron quiver velocity becomes marginally relativistic and the full relativistic treatment is necessary~\cite{brs}, which is ignored in this paper. 

The desired intensity for the coherent dynamical imaging~\cite{bio, bio2} is 
  $I=  10^{18} \ \mathrm{W} / \mathrm{cm}^2 $.
In Fig.~(\ref{fig:2}), the 1-D simulation of Eq.~(\ref{eq:2}) 
shows that the desired intensity for the coherent dynamical imaging can be reached 
without any additional focusing. 
In Fig.~(\ref{fig:2}), the pump (x-ray) propagetes from the right (left) to the left (right) and the x-ray pulse extracts energy from the pump pulse via the BRS, resuting the final energy of the x-ray pulse larger  than  its initial energy by a factor of $10^{5}$. 
In this example,  the pump laser has the intensity of  $I= 3.6 \times 10^{16} \ \mathrm{W} / \mathrm{cm}^2 $ ($I=  10^{15} \ \mathrm{W} / \mathrm{cm}^2 $) in the co-moving frame (the laboratory frame);   
the  intensity of the pump laser (x-ray pulse) is lower (higher)  by 36 times in the laboratory frame  than in the co-moving frame  due to the Doppler's effect. 
For the simulation shown in Fig.~(\ref{fig:2}), the final peak intensity of the x-ray pulse  in the co-moving frame  is larger by 10 times  than the intensity of the pump laser; the attained peak intensity of the x-ray pulse   is $I= 1.1 \times 10^{19} \ \mathrm{W} / \mathrm{cm}^2 $ in the laboratory frame.

In summary, the scheme achieves the soft x-ray light laser with the highest gain-per-length as shown in Table~\ref{tb}. The gain-per-length can be as high as $10^3 / \mathrm{cm}$ to $10^6 / \mathrm{cm}$, which is order magnitude higher than the current technologies~\cite{laser, laser1, gain},  and 
the peak x-ray intensity can be as high as or even higher than the pump laser.
The desired peak intensity of the coherent x-ray is  $I=  10^{18} \ \mathrm{W} / \mathrm{cm}^2 $ for the coherent dynamical imaging~\cite{bio, bio2}
can be reached under the proposed scheme
without any additional focusings.

 The comparison between  ND:YAG laser and the CO2 laser is made in the Table~\ref{tb}; the ND:YAG laser is  more advantageous in one-time intense x-ray burst but the CO2 laser would be better for the industry applications that need the high conversion efficiency and high-repetition rate.
 As demonstrated in the Raman compression of the visible-light laser~\cite{Fisch, malkin1, sonbackward, BBRS, BBRS2, BBRS3}, the BRS  is much stronger than the individual electron.  As an illustration, considering the electron plasma with the density $10^{20} / \mathrm{cm}^3 $ and the visible-light laser with the intensity of $10^{15}  \ \mathrm{W} / \mathrm{cm}^2$,  the BRS is $10^{9}$ times larger than the conventional Thomson scattering. This strong BRS is the main motivating physics that the author tries to utilize in this paper.

The performance of the proposed scheme relies on the quality of the electron beam such as the uniformity and the time duration in which the beam maintains its quality. Electron beam can deteriorate in a very fast time scale due to the space charge effect and in order to mitigate this effect, it probably needs to propagate inside a plasma of comparable electron density. The inverse bremsstrahlung is another concern. For such a high beam density, the inverse bremsstrahlung rate could be as high as $0.001  \ \omega_{pe}$ and higher for ultra-intense laser. If the pump laser is very intense, the heating by the inverse bremsstrahlung increases the electron temperature considerably in a few hundred Langmuir periods and the excitation of the appropriate Langmuir wave is no-longer feasible as  the condition given in Eq.~(\ref{eq:cond2}) is breached due to the heavy Landau damping.  
However, even if  the performances are compromised due to the mentioned issues,
   the scheme proposed  is important given the 
the possible highest gain strength (conversion efficiency). 

\bibliography{tera2}

\end{document}